\documentclass[useAMS,usenatbib]{mn2e}
\usepackage{txfonts}
\usepackage{amssymb}
\usepackage{epsf,graphicx,rotating}
\usepackage{color}
\usepackage[colorlinks,citecolor=blue]{hyperref}
\newcommand{\FeK}{Fe $\mathrm{K\alpha}$ }
\newcommand{\Xspec}{\textsc{Xspec}}
\newcommand{\apj}{ApJ}
\newcommand{\apjl}{ApJL}
\newcommand{\araa}{ARA\&A}
\newcommand{\aj}{AJ}
\newcommand{\apjs}{ApJS}
\newcommand{\aap}{A\&A}
\newcommand{\pasp}{PASP}
\newcommand{\pasj}{PASJ}
\newcommand{\mnras}{MNRAS}
\newcommand{\nat}{Nature}

\title[Relativistic Fe K$\alpha$ Line revealed in the composite X-ray spectrum of NLS1s]
{Relativistic Fe $\mathbf{K\alpha}$ Line Revealed in the Composite X-ray Spectrum of
Narrow Line Seyfert 1 Galaxies --- do their black holes have averagely low or intermediate spins?}
\author[Z. Liu, W.M. Yuan, Y.J. Lu \& X.L. Zhou] {Zhu Liu$^{1,2}$\thanks{E-mail:liuzhu@nao.cas.cn},
	 Weimin Yuan$^1$\thanks{E-mail:wmy@nao.cas.cn}, Youjun Lu$^1$ \& Xinlin Zhou$^1$\\
		 \\
$^1$ National Astronomical Observatories, Chinese Academy of Sciences, Beijing 100012, People's Republic of China\\
$^2$ University of Chinese Academy of Sciences, Beijing 100049, People's Republic of China\\ }
\date{Released \today}

\pagerange{\pageref{firstpage}--\pageref{lastpage}} \pubyear{2014}

\def\LaTeX{L\kern-.36em\raise.3ex\hbox{A}\kern-.15em
T\kern-.1667em\lower.7ex\hbox{E}\kern-.125emX}

\begin{document}

\label{firstpage}

\maketitle

\begin{abstract}

While a broad profile of the \FeK emission line is frequently found in the X-ray
spectra of typical Seyfert galaxies, the situation is unclear in the case of
Narrow Line Seyfert 1 galaxies (NLS1s)---an extreme subset which are generally
thought to harbor less massive black holes with higher accretion rates.  In
this paper, the ensemble property of the \FeK line in NLS1s is investigated by
stacking the X-ray spectra of a large sample of 51 NLS1s observed with {\it
XMM-Newton}. The composite X-ray spectrum reveals a prominent, broad emission
feature over 4--7$\,$keV, characteristic of the broad \FeK line. In addition,
there is an indication for a possible superimposing narrow (unresolved) line,
either emission or absorption, corresponding to Fe\,XXVI or Fe\,XXV,
respectively. The profile of the broad emission feature can well be fitted with
relativistic broad-line models, with the line energy consistent either with
6.4\,keV (i.e., neutral Fe) or with 6.67\,keV (i.e., highly ionized Fe), in
the case of the narrow line being emission and absorption, respectively.
Interestingly, there are tentative indications for low or
intermediate values of the average spins of the black holes ($a<0.84$),
as inferred from the profile of the composite broad line. If the observed
feature is indeed a broad line rather than resulting from partial covering
absorption, our results suggest that a relativistic Fe line may in fact be
common in NLS1s; and there are  tentative indications that black
holes in NLS1s may not spin very fast in general.
\end{abstract}

\begin{keywords} \vspace{1cm}
	X-ray:galaxies -- line:profile -- galaxies:Seyfert \end{keywords}


\section{Introduction}

The \FeK  emission  is the most important line feature in the X-ray spectra of
Seyfert galaxies \citep{Pounds+al+1990}.  Such a narrow line is commonly found
among Seyfert galaxies \citep{Nandra+Pounds+1994, Yaqoob+Padmanabhan+2004,
Zhou+Wang+2005}.  In addition, there is also observational evidence for a broad
\FeK line in some active galactic nuclei (AGN), whose archetypes include
MCG-6-30-15 \citep{Tanaka+al+1995, Fabian+al+2002, Miniutti+al+2007}, NGC 3516
\citep{Turner+al+2002, Markowitz+al+2006}, 1H0707-495 \citep{Fabian+al+2009}
and others \citep{Nandra+al+2007, Miller+2007}. The broad \FeK
	line is generally thought to originate from `cold' gas in the proximity of
	the supermassive black hole, such as the accretion disc, via the K-shell
fluorescence process, illuminated by X-rays from a primary X-ray source, such
as a postulated hot corona above the accretion disc \citep[][
		alternative explanations have also been suggested, however; see
\citealt{Miller+al+2008} for the absorption-origin model and
\citealt{Tatum+al+2012} for the Compton-thick wind model]{Fabian+al+1989}.  In
the presence of strong gravitational field, the line can be significantly
broadened, shifted and distorted due to the effects of both General and Special
Relativity, including  Doppler boosting, gravitational redshift and the
transverse Doppler effect.  The emergent line has a characteristic shape and
the exact profile is determined by the physical properties of the disc and the
black hole.  The energy of the red wing of the broad line, redshifted due
mainly to the gravitational redshift effect, is determined by the inner radius
of the accretion disc, which is commonly thought to be at the innermost stable
circular orbit (ISCO). The location of the ISCO is strongly dependent on the
spin of the black hole; it is at 6\,$r_{\rm g}$ ($r_{\rm g}=GM/c^2$, gravitational radius)
for a non-spinning black hole and 1.24\,$r_{\rm g}$ for a maximally spinning black
hole (dimensionless spin parameter $a=0.998$).  The energy and profile of the
broad line are also dependent on the ionization state and the inclination of
the accretion disc \citep{Ross+Fabian+1993, Fabian+al+2000}. As such, the
profile of a relativistic broad \FeK line can be used to constrain the spin of
the black hole \citep{Brenneman+Reynolds+2006, Dauser+al+2010,
Risaliti+al+2013}, as well as to probe the physical environment in the vicinity
of the black hole.

There have been extensive studies for characterizing the \FeK line properties
of Seyfert galaxies in the literature. For instance, \citet{Nandra+al+1997}
analyzed the spectra of 18 Seyfert 1 galaxies observed with {\it ASCA}, and
concluded that nearly 75\,per cent objects of their sample show evidence of
broad lines. Based on {\it XMM-Newton} spectra of a sample of 26 Seyfert
galaxies, \citet{Nandra+al+2007} found that almost all the spectra show a
narrow `core' at 6.4\,keV while $\sim$45 per cent of the spectra can be best
fitted with a relativistic blurred reflection model.
\citet{deLaCalle+al+2010} systematically and uniformly analyzed a
	collection of 149 radio-quiet Type 1 AGN observed by {\it XMM-Newton}. They
	found that about 36 per cent of sources in their flux-limited sample show
	strong evidence of a relativistic \FeK line with an average line equivalent
width ($EW$) of the order of 100\,eV, and this fraction can be considered as
a lower limit in the wider AGN population. By analysing the spectra of 46
objects observed with {\it Suzaku}, \citet{Patrick+al+2012} found that almost
all the objects reveal a narrow \FeK line and 50 per cent of the sample show a
statistically significant relativistic \FeK line with a mean equivalent width
$\sim$$96\pm10$\,eV.  It is generally accepted that a broad \FeK line is
frequently presented in Seyfert galaxies.

For individual objects, however, it requires a large number of X-ray photons
collected at high energies ($\gtrsim 4$\,keV) to determine accurately the
continuum spectrum and to reveal unambiguously the \FeK line profile
\citep{deLaCalle+al+2010}.  This is often unrealistic for the majority of AGN
given the limited observing time available in practice, except for a small
number of the brightest ones.  Alternatively, spectral stacking is an effective
way to obtain a composite spectrum with very high signal-to-noise for a certain
class of objects.  By stacking the X-ray spectra of 53 type 1 and 41 type 2
AGN, \citet{Streblyanska+al+2005} found broad relativistic \FeK lines with an
$EW$ of 600\,eV and 400\,eV for type 1 and type 2 AGN, respectively.
\citet{Corral+al+2008} introduced a sophisticated stacking method and applied
it to an AGN sample (mostly quasars)  observed with {\it XMM-Newton}; they
found a significant unresolved narrow \FeK line around $\sim$6.4\,keV with an
$EW\sim90$\,eV in the stacked spectrum, and only an indication for a broad \FeK
line which is not significant statistically. Using a sample of 248
AGN with a wide redshift range, \citet{Chaudhary+al+2012} analyzed the
average \FeK line profile using two fully independent rest-frame stacking
methods. They found that the average \FeK profile can be best fitted by a
combination of a narrow and a broad line with an $EW$ of $\sim$30\,eV and
$\sim$100\,eV respectively. \citet{Iwasawa+al+2012} performed spectral stacking
analysis of the XMM-COSMOS field to investigate the \FeK line properties of AGN
beyond redshift $z>1$; for type 1 AGN, they found a broad Fe emission line at a
low significant level ($\sim$2$\sigma$) for only sub-samples of low X-ray
luminosities ($<3\times10^{44}$\,ergs\,s$^{-1}$) or intermediate Eddington
ratios ($\lambda\sim0.1$); for type 2 AGN, no significant broad
Fe line is found.

For Narrow Line Seyfert 1 galaxies (NLS1s), the situation is far less clear,
however, given that the broad Fe line is detected at high significance in a few
objects only.  NLS1s are a subset of AGN defined as having the full width at
half maximum (FWHM) of their broad Balmer lines smaller than $\sim$$2000\rm{~km~s^{-1}}$
\citep{Osterbrock+Pogge+1985,Goodrich+1989}.  Compared to typical
Seyferts---broad-line Seyfert 1 galaxies (BLS1s), NLS1s generally show strong
$\mathrm{Fe\,{II}}$ and weak $\mathrm{[O\,III]}$ emission
\citep{Veron+al+2001}, steep soft X-ray spectra \citep{Puchnarewicz+al+1992,
Wang+al+1996, Grupe+2004}, and sometimes extreme X-ray variability
\citep{Leighly+1999, Grupe+2004}. It is generally accepted that NLS1s have
higher accretion rates close to the Eddington rate \citep{Boroson+Green+1992}
and relatively small black hole masses than BLS1s
\citep{Boroson+2002,Komossa+Xu+2007}. It has also been suggested that NLS1s
might be young AGN with fast growing black holes at an early stage of evolution
\citep[e.g.,][]{Mathur+2000}.  Their accretion process may proceed in a somewhat
different form from the standard thin disc, e.g., a slim disc as suggested to
dominate in the regime of high accretion rates close to the Eddington rate or
even higher \citep{Abramowicz+al+1988, Mineshige+al+2000}. NLS1s may represent
the extreme form of Seyfert activity and may provide a new insight into the
growth of massive black holes and accretion physics \citep{Komossa+2008}.

As far as the broad \FeK line is concerned, somewhat contradictory results have
been reported in the literature. There are a few well-studied NLS1s showing
apparently a broad \FeK emission line, including 1H 0707-495
\citep{Fabian+al+2009} and {\it SWIFT} J2127.4+5654 \citep{Miniutti+al+2009,
Marinucci+al+2014}. In an X-ray study of a large sample of AGN in
\citet[][see also
\citealt{Guainazzi+Bianchi+Dovciak+2006}]{deLaCalle+al+2010}, broad \FeK
lines were detected in 4 out of 30 ($\sim$13 per cent) NLS1s in the full
sample and in 2 out of 4 in the flux limited sample.
\citet{Ai+al+2011} studied the X-ray properties of a sample of 13 extreme
NLS1s that have very small broad-line widths ($<1200\rm{~km~s^{-1}}$) and found
evidence for significant \FeK lines in none of the objects.
\citet{Zhou+Zhang+2010} found that narrow \FeK emission lines in NLS1s are
systematically much weaker than those in BLS1s. It should be noted that in
previous studies of NLS1s, the X-ray spectra of individual objects are
mostly of low S/N ratios at energies around the \FeK line and higher,
partly due to the relatively steep continua above 2\,keV.  This results in
low photon counts, and hence poorly determined continuum and low S/N,
around the \FeK line region. It is not known whether a broad \FeK line is
commonly present in NLS1s as a population. Furthermore, the detection of
relativistic \FeK lines at high significance would render possible
diagnostics to the spin of black holes in NLS1s. In this paper, we
investigate the ensemble property of the \FeK line of NLS1s by stacking the
X-ray spectra of a large sample of NLS1s observed with {\it XMM-Newton}.
We use the cosmological parameters $H_0 =
70\,\mathrm{km\,s^{-1}\,Mpc^{-1}}$, $\Omega_\mathrm{M} = 0.27$ and
$\Omega_\Lambda = 0.73$. All quoted errors correspond to the 90 per cent
confidence level unless specified otherwise.

\begin{figure*}
	\begin{center}
		\includegraphics[width=0.8\textwidth]{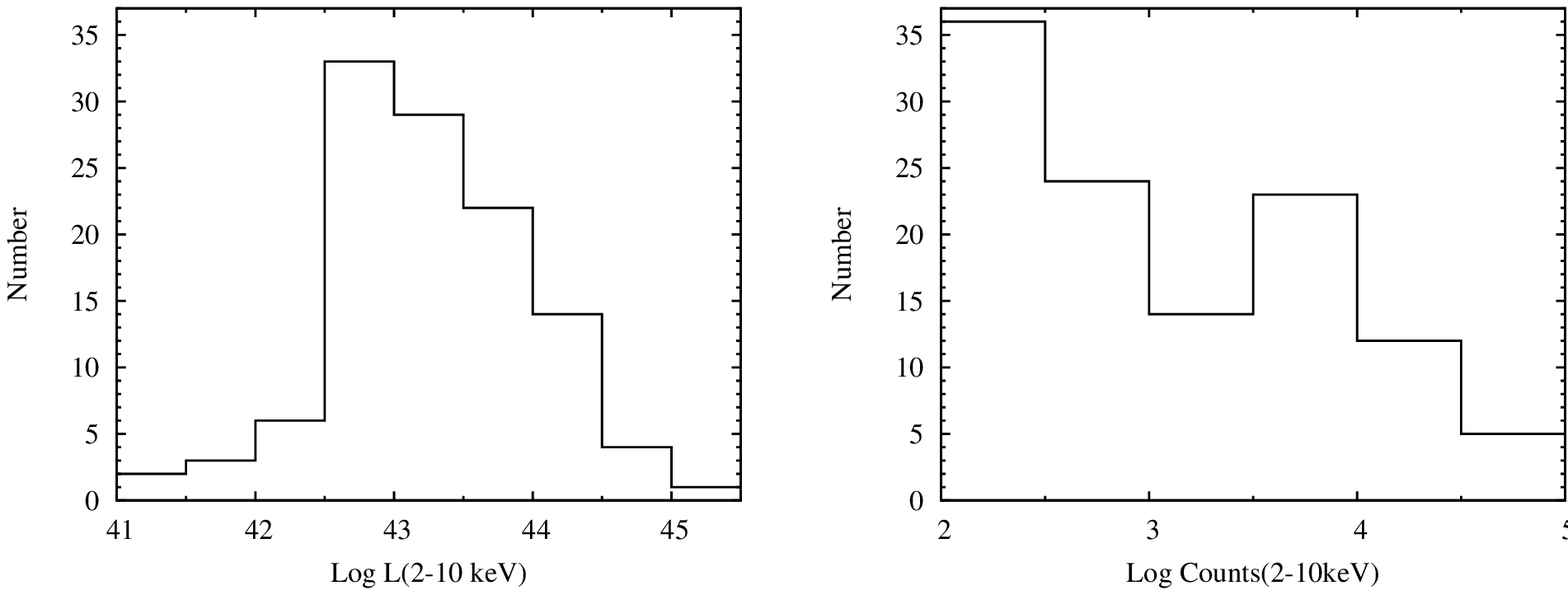}
		\caption{\label{fig:counts}{\it Left:} Distribution of 2-10\,keV rest-frame luminosity
		(MOS and PN) of the sample objects. {\it Right:} Distribution of
		2-10\,keV net photon counts.}
	\end{center}
\end{figure*}

\begin{table*}
\begin{turn}{90}
\begin{minipage}{235mm}
\begin{center}
    \caption{\label{tab:tab1}Basic parameters of the XMM-Newton sample}
\begin{tabular}{@{}lcccccccccccccc@{}}\\\hline\hline
Object name & R.A. & D.E.C & Obs\_ID & Redshift &\multicolumn{3}{c|}{EPIC MOS} && \multicolumn{3}{c|}{EPIC PN}  & N$_\mathrm{H}$ &FWHM(H$\beta$)&Ref.\\\cline{6-8}\cline{10-12}
			&J2000.0&J2000.0& & &Exposure & Counts Rate & $f_{2-10~\rm{keV}}$  &&Exposure & Counts Rate &$f_{2-10~\rm{keV}}$& & & \\
	(1)		& (2)  & (3) & (4) & (5) & \multicolumn{3}{c|}{(6)} && \multicolumn{3}{c|}{(7)} & (8) & (9) & (10)  \\\hline\hline
MARK  335			&   00 06 19.52 &+20 12 10.5 &0101040101&0.026 &33.68&1.197&1.43E-11&&28.47&1.631&1.33E-11&  3.56& $1710\pm140 $&  1  \\
                    &   \ldots      &\ldots      &0510010701&\ldots&20.02&0.203&3.41E-12&&15.56&0.290&2.98E-12&\ldots& \ldots       &  1  \\
I Zw 1              &   00 53 34.94 &+12 41 36.2 &0110890301&0.059 &  ---&  ---&    --- &&18.22&0.999&8.18E-12&  4.76& $1240		$&  2  \\
SDSS J010712+140844 &   01 07 12.04 &+14 08 45.0 &0305920101&0.077 &26.88&0.017&2.19E-13&&16.02&0.024&1.62E-13&  3.44& $787\pm31   $&  3  \\
MARK  359           &   01 27 32.55 &+19 10 43.8 &0112600601&0.017 &  ---&  ---&    --- &&6.84 &0.629&5.30E-12&  4.26& $900		$&  4  \\
RXS J01369-3510     &   01 36 54.40 &-35 09 52.0 &0303340101&0.289 &49.86&0.065&7.95E-13&&38.99&0.095&8.22E-13&  2.08& $1320       $&  5  \\
PHL 1092            &   01 39 55.75 &+06 19 22.5 &0110890901&0.396 &21.08&0.010&1.23E-13&&14.80&0.016&1.23E-13&  3.57& $1790		$&  2  \\
RX J0306.6+0003	    &   03 06 39.58 &+00 03 43.2 &0142610101&0.107 &52.03&0.058&8.21E-13&&40.30&0.081&7.94E-13&  6.31& $1904\pm51  $&  3  \\
RXS J03232-4931     &   03 23 15.35 &-49 31 06.7 &0140190101&0.071 &  ---&  ---&    --- &&24.90&0.153&1.30E-12&  1.34& $2075\pm70  $&  5  \\
RXS J04397-4540     &   04 39 44.84 &-45 40 42.0 &0204090101&0.224 &27.92&0.008&1.28E-13&&21.91&0.012&9.97E-14&  0.91& $2105\pm100 $&  5  \\
PKS 0558-504        &   05 59 47.38 &-50 26 52.4 &0116700301&0.137 &5.34 &0.668&8.11E-12&&  ---&  ---&    --- &  3.37& $1250		$&  6  \\
                    &   \ldots      &\ldots      &0117500201&\ldots&7.91 &1.064&1.26E-11&&  ---&  ---&    --- &\ldots& \ldots	     &  6  \\
                    &   \ldots      &\ldots      &0137550201&\ldots&14.02&0.800&9.90E-12&&10.59&1.191&1.04E-11&\ldots& \ldots	     &  6  \\
                    &   \ldots      &\ldots      &0137550601&\ldots&  ---&  ---&    --- &&10.22&2.436&2.10E-11&\ldots& \ldots	     &  6  \\
IRAS 06269-0543     &   06 29 24.67 &-05 45 29.6 &0153100601&0.117 &7.42 &0.175&2.23E-12&&2.59 &0.271&2.51E-12& 34.60& $\star$      &7,8 \\
SDSS J081053+280610	&   08 10 53.75 &+28 06 11.0 &0152530101&0.285 &27.02&0.004&4.66E-14&&16.21&0.007&6.46E-14&  2.89& $1212\pm78  $&  3  \\
SDSS J082433+380013	&   08 24 33.33 &+38 00 13.1 &0403760201&0.103 &20.35&0.005&5.78E-14&&14.00&0.009&5.79E-14&  4.06& $780\pm53	$&  3  \\
SDSS J082912+500652 &   08 29 12.80 &+50 06 52.0 &0303550901&0.043 &2.98 &0.051&4.75E-13&&  ---&  ---&    --- &  4.08& $603\pm21	$&  3  \\
SBS 0919+515        &   09 22 47.03 &+51 20 38.0 &0300910301&0.160 &21.07&0.027&3.37E-13&&10.17&0.033&2.58E-13&  1.33& $1132\pm27  $&  3  \\
MARK  110           &   09 25 12.87 &+52 17 10.5 &0201130501&0.035 &46.12&1.713&2.49E-11&&  ---&  ---&    --- &  1.30& $1760\pm50  $&  1  \\
SDSS 094057+032401 	&   09 40 57.20 &+03 24 01.2 &0306050201&0.061 &26.27&0.015&4.58E-13&&22.29&0.020&4.00E-13&  3.20& $1119\pm95  $&  3  \\
SDSS 094404+480646	&   09 44 04.41 &+48 06 46.6 &0201470101&0.392 &25.36&0.008&1.04E-13&&11.92&0.010&9.37E-14&  1.18& $1583\pm26  $&  3  \\
MARK 1239           &   09 52 19.10 &-01 36 43.5 &0065790101&0.020 &9.27 &0.015&5.52E-13&&  ---&  ---&    --- &  3.69& $1075		$&  4  \\
ACIS J10212+0421    &   10 23 48.44 &+04 05 53.7 &0108670101&0.099 &53.12&0.003&7.11E-14&&46.06&0.004&3.95E-14&  2.53& $696     $&9,10 \\
KUG 1031+398        &   10 34 38.60 &+39 38 28.2 &0109070101&0.042 &12.65&0.074&7.99E-13&&3.34 &0.102&8.23E-13&  1.31& $935		$&  4  \\
                    &   10 34 38.60 &+39 38 28.2 &0506440101&\ldots&87.28&0.076&8.23E-13&&72.67&0.107&8.15E-13&\ldots& \ldots       &  4  \\
SDSS 104613+525554	&   10 46 13.73 &+52 55 54.3 &0200480201&0.503 &18.37&0.010&1.25E-13&&7.24 &0.015&1.32E-13&  1.16& $2053\pm91  $&  3  \\
SDSS 112328+052823	&   11 23 28.12 &+05 28 23.3 &0083000301&0.101 &30.53&0.014&1.78E-13&&25.53&0.022&2.09E-13&  3.70& $1851\pm62  $&  3  \\
SDSS J11401+0307    &   11 40 08.71 &+03 07 11.4 &0305920201&0.081 &40.22&0.025&2.76E-13&&33.83&0.031&2.30E-13&  1.91& $675\pm41	$&  3  \\
PG 1211+143         &   12 14 17.66 &+14 03 13.3 &0112610101&0.081 &  ---&  ---&    --- &&48.84&0.325&2.78E-12&  2.74& $1860		$&  2  \\
                    &   \ldots      &\ldots      &0208020101&\ldots&41.07&0.258&3.30E-12&&32.94&0.350&3.05E-12&\ldots& \ldots       &  2  \\
                    &   \ldots      &\ldots      &0502050101&\ldots&49.48&0.358&4.37E-12&&44.07&0.468&3.88E-12&\ldots& \ldots       &  2  \\
                    &   \ldots      &\ldots      &0502050201&\ldots&33.00&0.379&4.71E-12&&25.71&0.507&4.20E-12&\ldots& \ldots       &  2  \\
SDSS 121948+054531	&   12 19 48.94 &+05 45 31.7 &0056340101&0.114 &29.56&0.011&2.17E-13&&  ---&  ---&    --- &  1.75& $1788\pm102 $&  3  \\
                    &   \ldots      &\ldots      &0502120101&\ldots&84.76&0.013&2.22E-13&&68.18&0.015&1.97E-13&\ldots& \ldots       &  3  \\
SDSS 123126+105111	&   12 31 26.45 &+10 51 11.4 &0306630101&0.304 &68.07&0.003&7.08E-14&&56.46&0.003&3.50E-14&  2.31& $957\pm23   $&  3  \\
                    &   \ldots      &\ldots      &0306630201&\ldots&91.88&0.003&7.03E-14&&81.51&0.003&4.89E-14&\ldots& \ldots       &  3  \\\hline\hline
\end{tabular}
\end{center}
\end{minipage}
\end{turn}
\end{table*}

\setcounter{table}{0}
\begin{table*}
\begin{turn}{90}
\begin{minipage}{235mm}
\begin{center}
\caption{Continue}
\begin{tabular}{@{}lcccccccccccccc@{}}\\\hline\hline
Object name & R.A. & D.E.C & Obs\_ID & Redshift &\multicolumn{3}{c|}{EPIC MOS} && \multicolumn{3}{c|}{EPIC PN}  & N$_\mathrm{H}$ &FWHM(H$\beta$)&Ref.\\\cline{6-8}\cline{10-12}
			&J2000.0&J2000.0& & &Exposure & Counts Rate & $f_{2-10~\rm{keV}}$  &&Exposure & Counts Rate &$f_{2-10~\rm{keV}}$& &  & \\
	(1)		& (2)  & (3) & (4) & (5) & \multicolumn{3}{c|}{(6)} && \multicolumn{3}{c|}{(7)} & (8) & (9) & (10)  \\\hline\hline
SDSS 123748+092323	&   12 37 48.50 &+09 23 23.2&0504100601& 0.125 &  ---&  ---&	--- &&17.63& 0.006& 5.24E-13&   1.48 & $968\pm74     $&  3  \\
WAS 61              &   12 42 10.60 &+33 17 02.6&0202180201& 0.044 &76.86&0.425&5.29E-12&&56.54& 0.568& 4.69E-12&   1.35 & $1900\pm150 $&  5  \\
                    &   \ldots      &\ldots     &0202180301& \ldots&11.71&0.356&4.57E-12&&9.16 & 0.480& 4.06E-12& \ldots & \ldots       &  5  \\
SDSS 124319+025256	&   12 43 19.98 &+02 52 56.2&0111190701& 0.087 &59.64&0.012&4.32E-13&&  ---&   ---&     --- &   1.92 & $1354\pm59  $&  3  \\
PG 1244+026         &   12 46 35.25 &+02 22 08.8&0051760101& 0.048 &  ---&  ---&    --- &&1.27 & 0.307& 2.32E-12&   1.87 & $740        $&  4  \\
SDSS 134351+000434	&   13 43 51.13 &+00 04 38.0&0202460101& 0.074 &26.05&0.006&5.11E-13&&20.47& 0.006& 1.22E-13&   1.89 & $2106\pm73  $&  3  \\
SDSS J134452+000520 &   13 44 52.91 &+00 05 20.2&0111281501& 0.087 &6.95 &0.015&2.34E-13&&  ---&   ---&     --- &   1.89 & $1997\pm38  $&  3  \\
IRAS 13224-3809     &   13 25 19.38 &-38 24 52.7&0110890101& 0.066 &61.22&0.044&3.53E-12&&51.40& 0.063& 5.01E-13&   5.34 & $650		  $&  2  \\
SDSS 133141-015212	&   13 31 41.03 &-01 52 12.5&0112240301& 0.145 &32.47&0.007&3.45E-12&&24.40& 0.009& 2.09E-13&   2.36 & $1192\pm42  $&  3  \\
IRAS 13349+2438     &   13 37 18.73 &+24 23 03.4&0402080201& 0.108 &32.83&0.288&2.30E-12&&  ---&   ---&     --- &   1.02 & $2200       $&  11  \\
                    &   \ldots      &\ldots     &0402080301& \ldots&58.58&0.275&6.27E-14&&  ---&   ---&     --- & \ldots & \ldots       &  11  \\
                    &   \ldots      &\ldots     &0402080801& \ldots&10.19&0.177&5.10E-13&&  ---&   ---&     --- & \ldots & \ldots       &  11  \\
2E 1346+2646        &   13 48 34.95 &+26 31 09.9&0097820101& 0.059 &43.13&0.061&9.85E-13&&34.34& 0.085& 8.69E-13&   1.18 & $1235^{a}   $&  4  \\
                    &   \ldots      &\ldots     &0109070201& \ldots&54.76&0.035&7.11E-13&&48.79& 0.016& 4.12E-13& \ldots & \ldots 	   &  4  \\
SDSS J13574+6525    &   13 57 24.51 &+65 25 05.9&0305920301& 0.106 &20.69&0.020&2.20E-13&&15.94& 0.030& 2.54E-13&   1.36 & $737\pm41   $&  3  \\
                    &   \ldots      &\ldots     &0305920601& \ldots&14.83&0.020&2.49E-13&&11.99& 0.033& 2.73E-13& \ldots & \ldots       &  3  \\
UM 650              &   14 15 19.50 &-00 30 21.6&0145480101& 0.135 &18.81&0.007&2.12E-13&&13.66& 0.009& 1.06E-13&   3.25 & $1045\pm27  $&  3  \\
Zw 047.107          &   14 34 50.63 &+03 38 42.6&0305920401& 0.029 &21.44&0.012&1.65E-13&&17.34& 0.016& 1.39E-13&   2.43 & $1089^{a}   $& 12  \\
MARK  478           &   14 42 07.46 &+35 26 22.9&0107660201& 0.079 &24.67&0.240&2.72E-12&&13.55& 0.316& 2.55E-12&   1.05 & $1270       $&  4  \\
PG 1448+273         &   14 51 08.76 &+27 09 26.9&0152660101& 0.065 &21.09&0.179&2.08E-12&&17.91& 0.242& 1.94E-12&   2.78 & $1050       $&  4  \\
IRAS 15091-2107     &   15 11 59.80 &-21 19 01.7&0300240201& 0.045 &12.88&0.614&8.93E-12&&5.02 & 0.877& 7.95E-12&   8.32 & $1480       $&  13  \\
SDSS 154530+484609	&   15 45 30.24 &+48 46 09.1&0153220401& 0.400 &7.97 &0.016&1.86E-13&&  ---&   ---&     --- &   1.64 & $1821\pm24  $&  3  \\
                    &   \ldots      &\ldots     &0505050201& \ldots&6.13 &0.028&3.75E-13&&3.28 & 0.038& 3.26E-13& \ldots & \ldots       &  3  \\
                    &   \ldots      &\ldots     &0505050701& \ldots&6.12 &0.022&2.35E-13&&  ---&   ---&     --- & \ldots & \ldots       &  3  \\
MARK  493           &   15 59 09.63 &+35 01 47.5&0112600801& 0.031 &  ---&  ---&    --- &&8.43 & 0.447& 3.66E-12&   2.11 & $740        $&  4  \\
KAZ 163             &   17 46 59.13 &+68 36 27.8&0300910701& 0.063 &11.66&0.096&2.20E-12&&7.69 & 0.115& 2.23E-12&   4.06 & $1600       $& 14  \\
PKS 2004-447        &   20 07 55.18 &-44 34 44.3&0200360201& 0.240 &34.33&0.070&9.44E-13&&25.17& 0.094& 8.73E-13&   3.17 & $1447       $&  15 \\
MARK  896           &   20 46 20.87 &-02 48 45.3&0112600501& 0.026 &10.44&0.321&3.91E-12&&7.19 & 0.426& 3.64E-12&   3.39 & $1135       $&  4  \\
XMM J20584-4236     &   20 58 29.91 &-42 36 34.3&0081340401& 0.232 &14.60&0.025&2.92E-13&&7.19 & 0.037& 3.09E-13&   3.31 & $\star$      &  16 \\
II Zw 177           &   22 19 18.53 &+12 07 53.2&0103861201& 0.082 &  ---&  ---&    --- &&7.96 & 0.121& 9.01E-13&   4.90 & $2080^{a}  $&17,18 \\
HB89 2233+134   	&   22 36 07.68 &+13 43 55.3&0153220601& 0.326 &8.68 &0.044&6.18E-13&&6.35 & 0.057& 4.83E-13&   4.51 & $1840\pm36  $&  3 \\\hline\hline
\end{tabular}

\parbox[]{21.5cm}{The columns are: (1) object name; (2) and (3)
	position in J2000; (4):observation ID; (5) redshift;
	(6)(7) exposure (in units of kiloseconds), 2-10\,keV counts rate and flux for EPIC MOS
	and PN; (8) Galactic column density; (9)FWHM of H$\beta$; (10) reference for the FWHM of H$\beta$,
  where 1: \citet{Grupe+al+2004}, 2: \citet{Leighly+1999}, 3:
  \citet{Zhou+al+2006}, 4: \citet{Veron+al+2001}, 5: \citet{Grupe+al+1999}, 6: \citet{Corbin+1997},
  7: \citet{Veron+Veron+2006}, 8: \citet{Moran+Halpern+Helfand+1996}, 9: \citet{Hornschemeier+al+2005},
  10: \citet{Greene+Ho+2007}, 11: \citet{Grupe+al+1998}, 12: \citet{Greene+Ho+2004}, 13:\citet{Boller+al+1996},
  14: \citet{Haddad+Vanderriest+1991}, 15: \citet{Oshlack+al+2001},
  16: \citet{Caccianiga+al+2004}, 17: \citet{Appenzeller+al+1998}, 
  18: \citet{Xu+al+2003}; (10) 2XMMcata-DR3 name

  ---: no data

  $a$: measured using broad H$\alpha$ line.

  $\star$: identified as NLS1s in the reference without giving the detail of the FWHM(H$\beta$)}
\end{center}
\end{minipage}
\end{turn}
\end{table*}


\section{Sample}

We compiled a sample of NLS1s that have {\it XMM-Newton} observations from two
NLS1s catalogues. The first by far is the largest NLS1s sample ($\sim$2000)
selected from the SDSS DR3 by \citet{Zhou+al+2006}, which is homogeneously
selected with well measured optical spectrometric parameters. The second
includes NLS1s from the AGN catalogue compiled by \citet{Veron+Veron+2006}
comprising about 400 objects collected from the literature. We searched for {\it XMM}
observational data from the {\it XMM} archive that are included in the  Second {\it
XMM-Newton} Serendipitous Source Catalogue
\citep[2XMMi-DR3,][]{Watson+al+2009}. We selected those having more than
100 net source counts detected with either the EPIC PN or the
combined MOS detectors in the 2-10\,keV band.  Three objects are excluded that
are previously known to show a significant broad \FeK line, namely, 1H
0707-495, Mrk 766, and NGC 4051\footnote{It should be noted that NGC 4051 has a
	low Eddington ratio, $\lambda\le0.03$ \citet{Peterson+al+2004,
Denney+al+2009}, and we thus  do not consider it to be a typical NLS1s, but
rather a normal Seyfert galaxy with a small-mass  black hole (and therefore a
narrow line width; see \citealt{Ai+al+2011} for discussion).}. The sample
includes 51 NLS1s that have a total of 68 observations. A small number
of sources have multiple observations; for them no significant variability is found,
and each of their observations is treated as an independent spectrum. The basic parameters
of the sample are presented in Table \ref{tab:tab1}. The redshifts and
FWHM(H$\beta$) are taken from \citet{Zhou+al+2006} if available, otherwise from
the NASA Extragalactic Database (NED) or from the literature. The sample
objects have redshifts $\leq0.5$, with a median of $\sim0.085$. A small number
of sources claimed to be NLS1s but having no H$\beta$ line width values are
denoted by $\star$. The distribution of the H$\beta$ FWHM has a median of
$\sim$1330$~\mathrm{km~s^{-1}}$.

The X-ray spectra are retrieved from the LEDAS
website\footnote{\url{http://www.ledas.ac.uk/}}, which were reduced by the XMM
pipe-line using the standard algorithm \citep{Watson+al+2009}. For each
observation the MOS1 and MOS2 spectra are combined into one single MOS
spectrum. This results in a total of 114 X-ray spectra, with the PN and MOS
spectra treated independently. The distributions of
luminosity and the net photon count in the 2--10\,keV band
are shown in Figure \ref{fig:counts}. The median of the rest frame 2--10\,keV luminosity is
$1.92\times10^{43}$\,ergs\,s$^{-1}$ and about 83 per cent of the sample
have luminosities below $10^{44}$\,ergs\,s$^{-1}$.  Half of the sample have
counts less than 1000 and 85 per cent have counts less than 10000.

\section{Composite X-ray Spectrum of NLS1s}
\subsection{Spectral stacking}

In X-ray observations, the measured spectra are a result of the convolution
of the source spectra, usually  modified by  absorption, with the
instrumental responses, which are functions of   photon energy and
vary with the source positions across the detectors.
In general, the measured spectra of
different sources/observations and instruments cannot be co-added directly.
This is especially true for  sources observed at different redshifts.
In this work, we adopt a
method introduced by \citet{Corral+al+2008} to stack the X-ray spectra of
objects with different redshifts from different observations. We refer readers
to the original paper for details and discussions about the method, and
only outline the main procedures here.
\begin{enumerate}
    \item Determining the source continuum: Each of the observed spectrum is
        fitted with a simple power-law modified by Galactic and possible
        intrinsic absorption ({\sc po*pha*zpha} in \Xspec,
        \citealt{Arnaud+1996}). Only the spectrum above 2\,keV is used to
        avoid possible contamination from the soft X-ray excess commonly found
        in the X-ray spectrum of AGN below 2\,keV. The 5--7\,keV energy range
        is excluded to exclude possible contribution from potential \FeK
        emission line features. The Galactic absorption column density is fixed
        at the Galactic value given by \citet{Kalberla+al+2005} for each
        source. The intrinsic absorption column density, the slope and
        normalization of the power law are left as free parameters.  We find
        that the above model can  describe the continuum spectra very well for
        all the sample objects.  In this way, the best fitted continuum model
        is obtained. We then include the 5--7\,keV energy range before we unfold
        (see below) the X-ray spectra.
    \item Unfolding the spectrum: The intrinsic source spectra (before entering
        the X-ray telescope) can be reconstructed with the instrumental effects
        eliminated by unfolding the observed spectra with the calculated
        conversion factors from the best-fit models. We apply the best-fit
        continuum model to the ungrouped spectra. The unfolded spectra are then
        calculated by using the {\tt eufspec} command in {\Xspec }.
    \item Rescaling: The unfolded spectra are corrected for the absorption
        effects (both Galactic and intrinsic) and de-redshifted to the source
        rest frame. We rescale the normalization of the unfolded spectra in
        such a way that each spectrum has nearly the same weight as the others;
        following \citet{Corral+al+2008}, we rescale the spectra in such a way
        that they have the same 2--5\,keV fluxes in the rest frame.
    \item Rebin: The de-redshifted and rescaled rest-frame spectra have
        different energy bins (as they are set by the energy channels of the
        detectors), and thus have to be rebined into unified bins before
        stacking. To construct a new, unified bin scale, we shift the observed
        spectra (the original folded and unrescaled spectra, measured in counts
        per channel) to the source rest frame, rebin the spectra to a common
        energy grid comprised of narrow bins (bin width $\sim$50\,eV). We add
        all the rebinned spectra together in counts per bin.  The co-added
        spectrum is grouped in such a way that each new bin contains at least
        900 counts. This sets the final bin scale for the stacking. Finally, we
        rebin the unfolded and rescaled spectra according to this new energy
        bin system; In this way,  all the unfolded and rescaled spectra have
        the same energy bins.
	\item Stacking: The stacked spectrum is obtained by averaging the rebinned
		and rescaled spectra using the simple arithmetic mean. The errors are
		calculated using the propagation of errors, as in \citet{Corral+al+2008}.
\end{enumerate}

\begin{figure}
	\begin{center}
	\includegraphics[width=0.9\columnwidth]{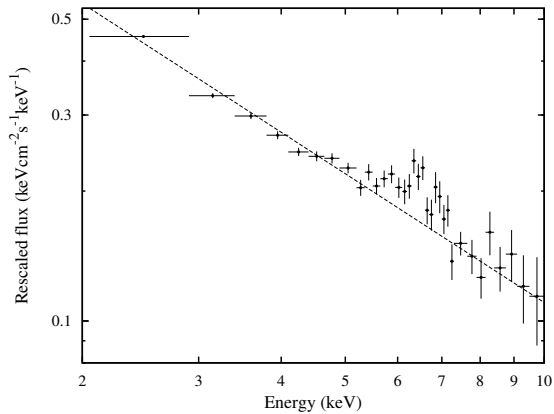}
	\caption{The stacked X-ray spectrum of our NLS1s sample;
	the dashed line represents the best-fit of the continuum with a  power-law model.}
	\label{fig:aver_fit}
	\end{center}
\end{figure}

\subsection{Fe K{\boldmath$\alpha$} emission line feature and its reliability}
\label{sec:subsec3.2}

The stacked 2--10\,keV spectrum is shown in Figure \ref{fig:aver_fit}. Fitting
the continuum spectrum with the 5--8\,keV range excluded gives a photon index
$\Gamma=1.99\pm0.04$.  It shows clearly a prominent broad emission feature in
the 5--7\,keV energy range superposing the power-law continuum, which must be
due to the \FeK emission.  In below we test the reliability of the presence of
this feature, by means of Monte Carlo simulations.
		
\begin{figure*}
	\begin{center}
	\includegraphics[width=0.8\textwidth]{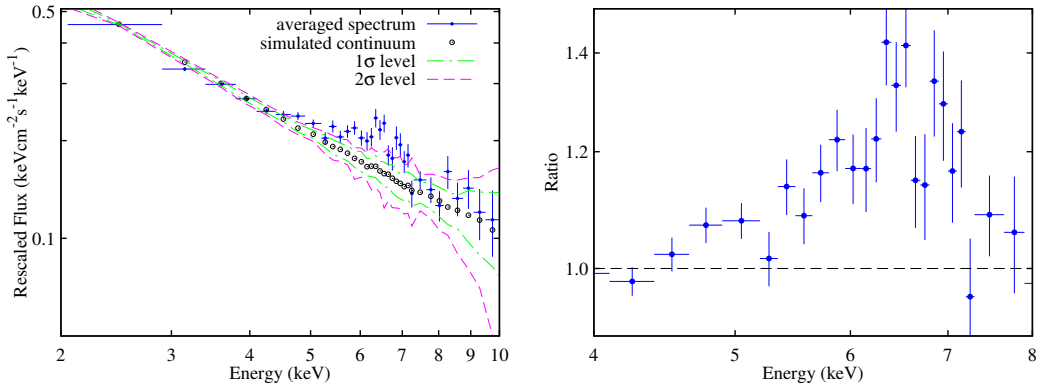}
	\caption{{\it Left-hand panel:} Stacked source spectrum (blue points with
		error bars), simulated power-law continuum (open circles), and its
		confidence ranges at the 1$\sigma$ (68 per cent) and 2$\sigma$ (90 per
		cent) levels for each energy bin, respectively.  The green dot-dashed
		line and the red dashed line are the 1$\sigma$ and 2$\sigma$ levels for
	each bin, respectively.  {\it Right-hand panel:} Ratio of the stacked
source spectrum to the stacked simulated power-law continuum in the 4--8\,keV
range,  resembling closely the characteristic relativistic broad \FeK line.}
\label{fig:aver_con_sig}
	\end{center}
\end{figure*}

In their original work, \citet{Corral+al+2008}  performed extensive simulations
to investigate the effects of the stacking method on the detection of potential
line features in the stacked spectrum, and concluded that the method is
reliable and does not  produce spurious line features nor change the profile of
real lines.  Here we carry out independent checks to test whether the broad
emission feature is an artifact produced by the stacking procedure, using
simulations similar to those in \citet{Corral+al+2008}. This is done by
applying the stacking method to a set of simulated `observed' power-law spectra
without any lines.  Using the best-fit continuum spectrum model (power-law plus
absorption), we simulate the `observed' source and background spectra for each
of our objects using {\Xspec}; the exposure time, auxiliary and the response
matrix files for the sources are used  in the simulations to match exactly the
observations.  We then apply the same stacking method to those simulated
spectra and obtain a composite simulated spectrum, which is shown in Figure
\ref{fig:aver_con_sig} (left-hand panel; open circles), along with the
composite source spectrum for a comparison. As expected, no emission feature is
present in the stacked simulated continuum. Thus the broad emission feature in
composite source spectrum is not an artifact caused by the stacking method.  A
power-law gives a good fit to the composite simulated continuum with
$\Gamma=2.05\pm0.01$, well consistent with the model fitted to the underlying
continuum of the stacked source spectrum above. Following
\citet{Corral+al+2008}, we use this composite simulated continuum as the
underlying continuum of the composite source spectrum.

To construct the confidence intervals for the underlying continuum of the
composite source spectrum, we simulate 100 power-law plus absorption continuum
spectra for each individual source and produce 100 stacked continuum spectra.
By removing the far-most, two-sided 32 and 5 extreme values in each energy bin,
we can estimate the 68 per cent (1$\sigma$) and 95 per cent (2$\sigma$)
confidence intervals for the underlying continuum of the stacked spectrum,
which are over-plotted in Figure \ref{fig:aver_con_sig} (left panel).  For a
given energy bin, the probability that the data point of a simulated stacked
spectrum falls out of the 2$\sigma$ interval is $p<0.05$. The joint
probability that $N$ adjacent data bins are all in excess of the 2$\sigma$
confidence region is $P<(0.05)^N$. It can be seen that at least 10 adjacent
data points of the composite source spectrum in the \FeK region fall at or out
of the 2$\sigma$ confidence levels. Therefore, the probability that the
observed broad emission feature in the composite source spectrum arises from
statistical fluctuations is $P<5\times 10^{-20}$.

Furthermore, we investigate whether the  profile of a line can be significantly
affected by the stacking method. We simulate a series of observed spectra, with
the intrinsic spectra comprising a power law continuum and a Gaussian line,
using the response matrices and exposures in our sample.  The line energy and
the line width $\sigma$ are fixed at reasonable values (e.g., $E=6.4$\,keV,
$\sigma=100$\,eV or less), and the other parameters are randomly chosen (e.g.,
$\Gamma$ in the 2.0--2.2 range and the $EW$ in a wide range). We then apply the
stacking method to the simulated spectra. We find that the method has almost no
affect on the line profile if the line is significantly broader than the
instrumental energy resolution, as in our case here. However, as
expected, narrow lines remain unresolved in the unfolded spectra with a
width comparable to the intrinsic instrumental energy resolution, which is
energy dependent \citep{Corral+al+2008, Iwasawa+al+2012}. We examine the resulting
profile in the unfolded spectra of an intrinsically unresolved Gaussian line
(assuming $\sigma=1$\,eV) by simulations. We consider the line energy in the
Fe\,K$\alpha$ line region 6.4--6.97\,keV (Fe\,I--Fe\,XXVI) and a redshift of
0.085 (the median redshift of our sample). We find that an intrinsically unresolved
line manifests itself as a Gaussian in the unfolded spectrum, which has
a standard deviation around $\sigma\sim85$\,eV
(70--110\,eV). These results are consistent with the conclusion reached by
\citet{Corral+al+2008}.

We thus conclude that the broad emission feature in 5-7\,keV in the composite
source spectrum should be real, rather than being an artifact produced by the
stacking method nor due to statistical fluctuations. It should be noted that 
this stacking method is based on essentially averaging of the fluxes of the
sources rather than photon counts, and the way of weighting is such that all
the spectra have nearly the same weight by rescaling (see above). As a result,
the composite spectrum will not be dominated by strong sources\footnote{As a
demonstration, we divide the sample into two sub-samples, one comprising of
sources with net counts $<5000$ and the other with net counts $>5000$. We
find that the overall profiles of the stacked spectra of the two sub-samples are
well consistent with each other.}.
    
The right-hand panel in
Figure \ref{fig:aver_con_sig} shows the ratio of the stacked source spectrum to
the stacked simulated power-law continuum. It can be seen that its profile
resembles closely a relativistic broad \FeK line. In section \ref{sec:sec4} we
study the profile of this complex feature by modelling it with relativistically
broadened line models as well as other models.

\section{Modeling the broad line feature}
\label{sec:sec4}

We use \Xspec\, (version 12.8) to fit the stacked spectrum. The {\tt FTOOLS} task
{\tt flx2xsp} is used to convert the flux spectrum into fits format which can
be fitted using \Xspec.
Since we are interested in the broad emission feature, we use only the
3--8\,keV band in the spectral fitting below. This avoids the somewhat large
uncertainty in the higher energy band and the highly model-dependent unfolded
spectrum (e.g., absorption, soft X-ray excess emission) below 3\,keV
\citep{Corral+al+2008}. A simple power-law model leads to an unacceptable fit
($\mathrm{\chi^2/d.o.f}=156/22$), as expected, and other components have to be
added. In the following fittings, we adopt the power-law fitted from the
stacked simulated spectrum as the continuum model, with the slope and
normalization parameters fixed at the best-fit values obtained above. In
addition, a Gaussian component\footnote{\label{fn:lw}In the actual
fitting the width of the Gaussian line is fixed at $\sigma=85$\,eV to
account for the unresolvability of narrow lines in the unfolded spectra
due to the intrinsic instrumental energy resolution (see
Section\,\ref{sec:subsec3.2}; using other values of 70--110\,eV has almost no
affect on the results). The same practice is exercised for adding other unresolved
narrow lines in the modelling in the rest of the paper.} with the line energy
fixed at 6.4\,keV is also included to account for the unresolved narrow \FeK
line which is ubiquitously found in AGN. We
consider this model of the power-law plus the narrow Gaussian line as a
baseline model. This baseline model only results a poor fit
($\mathrm{\chi^2/d.o.f}=106/25$), with an $EW\sim114$\,eV of the narrow line.
Adding another Gaussian component with all the parameters free can improve the
fit significantly ($\mathrm{\chi^2/d.o.f}=21/22$). This broad Gaussian
component, with the line energy peaked at $\sim$6.2\,keV and line width
$\sim$0.8\,keV, may indicate a broad line feature in the $\sim$5--7\,keV energy
range.  Below we try to fit this broad profile using various models.

\begin{table*}
	\begin{minipage}{170mm}
	\begin{center} \caption{Spectral fit in the 3.0-8.0\,keV energy range} \label{tab:table3}
        \begin{tabular}{@{\extracolsep{4pt}}ccccccccccc@{}}\\\hline
	\multicolumn{9}{l}{\textbf{Baseline Model: Power-Law + Gaussian}}\\\hline
    {\sc Power-Law} & \multicolumn{3}{c}{\sc Gaussian}        & & && & $\chi^2/\rm{d.o.f}$ \\\cline{1-1}\cline{2-4}
    $\Gamma$        & $E$(keV)   & $\sigma~$(keV) & $EW~$(eV) & & && & \\
    2.05(fixed)     & 6.4(fixed) & 0.085$^{c}$(fixed)   & $\sim$114  & & && & 106/25\\\hline\\[0.2mm]
	\multicolumn{9}{l}{\textbf{Baseline$^{\star}$ + Partial Covering}}\\\hline
    \multicolumn{2}{c}{\sc Baseline}                      && \multicolumn{2}{c}{\sc Pcfabs}            &&&&$\chi^2/\rm{d.o.f}$ \\\cline{1-2}\cline{4-5}
    $\Gamma$                    & $EW_{6.4~\rm{keV}}$(eV) &&$N_{\rm H}$$^{a}$ & covering factor        &&&&\\
    $\rm{2.31^{+0.27}_{-0.26}}$ & $\rm{43\pm{25}}$        && $\rm{44^{+20}_{-12}}$ &$\rm{0.48^{+0.12}_{-0.15}}$ &&&&18/21\\\hline\\[0.2mm]
    \multicolumn{9}{l}{\textbf{Baseline + LAOR + Emission Line(6.97~keV)/Absorption Line(6.67~keV)}}\\\hline
    \multicolumn{2}{c}{\sc Baseline}& \multicolumn{3}{c}{\sc laor}                                            &\multicolumn{3}{c}{\sc Gaussian}              & $\chi^2/\rm{d.o.f}$ \\\cline{1-2}\cline{3-5}\cline{6-8}
	$\Gamma$    &$EW_{6.4~\rm{keV}}$(eV) & $E$(keV)               & $R_{\rm in}$                  & $EW~$(eV)      &$E$(keV)   & $\sigma~$(keV) & $EW~$(eV)       & \\
    2.05(fixed) &$    27^{+30}_{-27}$      & $6.38^{+0.45}_{-0.08}$ & $4.4^{+5.3}_{-1.7}~R_{\rm g}$ & $\rm{362\pm90}$&6.97(fixed)&0.085$^{c}$(fixed)    &$\rm{73\pm{30}}$ &23/21 \\
    2.05(fixed) &$\rm{33\pm{27}}$        & $6.75\pm{0.10}$& $3.7^{+1.8}_{-1.2}~R_{\rm g}$&$\rm{580\pm120}$&6.67(fixed)& 0.085$^{c}$(fixed)   & $    20^{+22}_{-20}$  & 21/21 \\\hline\\[0.2mm]
    \multicolumn{9}{l}{\textbf{Baseline + RELLINE + Emission Line(6.97~keV)/Absoprtion Line(6.67~keV)}}\\\hline
    \multicolumn{2}{c}{\sc Baseline}      & \multicolumn{3}{c}{\sc relline}                       & \multicolumn{3}{c}{\sc Gaussian}                  & $\chi^2/\rm{d.o.f}$ \\\cline{1-2}\cline{3-5}\cline{6-8}
	$\Gamma$    & $EW_{6.4~\rm{keV}}$(eV) & $E$(keV)      &     spin          & $EW~$(eV)         & $E$(keV)   & $\sigma~$(keV) & $EW~$(eV)           & \\
    2.05(fixed) & $    21^{+30}_{-21}$    & $6.34\pm0.04$ & $<0.66$& $\rm{364\pm87}$   & 6.97(fixed)&0.085$^{c}$(fixed)    & $\rm{76\pm{30}}$    &23/21 \\
    2.05(fixed) & $\rm{33\pm{27}     }$   & $6.73\pm0.04$ & $0.47^{+0.34}_{-0.39}$ & $\rm{570\pm120}$  & 6.67(fixed)& 0.085$^{c}$(fixed)   & $    21^{+22}_{-21}$  & 23/21 \\\hline\\[0.2mm]
    \multicolumn{9}{l}{\textbf{Baseline$^{\star}$ + RELLINE + Emission Line(6.97~keV)/Absorption Line(6.67~keV)}}\\\hline
    \multicolumn{2}{c}{\sc Baseline}          & \multicolumn{3}{c}{\sc relline}                &\multicolumn{3}{c}{\sc Gaussian}                   & $\chi^2/\rm{d.o.f}$ \\\cline{1-2}\cline{3-5}\cline{6-8}
    $\Gamma$             & $EW_{\rm{NL6.4~keV}}$(eV)&$E$(keV)       &    spin   &       $EW~$(eV)    & $E$(keV)   & $\sigma~$(keV) & $EW~$(eV)           & \\
    $\rm{1.92\pm{0.08}}$ & $    13^{+39}_{-13}$ &$6.33\pm{0.06}$& $<0.40^b$ & $\rm{233\pm{113}}$ & 6.97(fixed)&0.085$^{c}$(fixed)    &$\rm{49\pm{32}}$     &17/19 \\
    $\rm{1.92\pm{0.09}}$ & $\rm{34\pm{27}}$  & $6.71\pm{0.09}$& $<0.84^b$ & $\rm{400\pm{227}}$ & 6.67(fixed)& 0.085$^{c}$(fixed)   & $    18^{+23}_{-18}$ & 18/19 \\\hline
	\end{tabular}
	\parbox[]{17.0cm}{
		$\star$: In this model, the photon index and normalization of the power-law are free parameters.\\
		$a$: In unit of $10^{22}$~cm$^{-2}$.\\
        $b$: $1\sigma$ confidence level.\\
        $c$: This width corresponds to an unresolved narrow line in the `intrinsic' spectra; see footnote~\ref{fn:lw} and
        Section~\ref{sec:subsec3.2}
	}
	\end{center}
	\end{minipage}
\end{table*}

\subsection{\label{sec:subsec4}Relativistic broad line models}

The broad feature around 6.4\,keV observed in many Seyfert galaxies is widely
believed to arise from the \FeK line emitted from the inner accretion disk close
to the black hole, which is highly broadened and skewed due to relativistic
effects and gravitational redshift. In this section we fit this feature in the
stacked spectrum with such a relativistic line.

\textbf{\textit{Laor model:}} We first adopt the {\sc laor} model
\citep{Laor+1991}, which has been widely used in fitting relativistic \FeK
lines in AGN.  The {\sc laor} model is fully relativistic, assuming the Kerr
metric for a black hole with the maximum spin (dimensionless spin parameter
$a=0.998$). The line energy, inner radius ($R_\mathrm{in}$) of the accretion
disc and the normalization are free parameters, while the other parameters,
which can not be well constrained, are fixed at their default values, e.g., the
disc inclination $i=30$\,degree and the emissivity index $q=3.0$.
The {\sc laor} model leads to a poor fit
($\mathrm{\chi^2/d.o.f}=41/22$) with the best-fit line energy at
$\sim$6.39\,keV. The large $\chi^2$ value results from a systematic
structure in the residuals, which appears to be either an absorption due to
Fe\,XXV (6.67keV) or an emission line identified as Fe\,XXVI (6.97keV).  We
consider these two scenarios respectively, by adding to the model an
unresolved narrow Gaussian as an emission component (referred to as Case A
hereafter) and as an absorption component (Case B); the line energies are
fixed at the corresponding values and the line-widths are fixed at 85 eV
(corresponding to an unresolved narrow line in the intrinsic spectrum; see
footnote\,\ref{fn:lw} and Section\,\ref{sec:subsec3.2}). As can be seen
below, this improves the fitting significantly and results in acceptable fits
(Table\,\ref{tab:table3}).

For Case A (narrow emission line of Fe\,XXVI), the best-fit line
energy of the broad line component is $6.38^{+0.45}_{-0.08}$\,keV with an $EW$
of $362\pm{90}$\,eV. The best-fit inner radius of the disc is $R_{\mathrm{in}}=
4.4^{+5.3}_{-1.7}~R_{\rm g}$. The $EW$s of the neutral narrow \FeK line and
the high ionization narrow line components are 27\,eV with an upper
limit of 57\,eV and $73\pm30$eV, respectively. The fitted parameters are
given in Table\,\ref{tab:table3}.
Alternatively, for Case B (narrow absorption line of Fe\,XXV), the
line energy of the broad component is $6.75\pm{0.10}$\,keV, which may correspond
to the K$\alpha$ transition of Fe\,XXV ions. The best-fit inner radius of the
disc is $R_{\mathrm{in}}= 3.7^{+1.8}_{-1.2}~R_{\rm g}$. The $EW$s of the broad
component, the narrow \FeK component and the narrow absorption component are
$580\pm{120}$\,eV, $33\pm{27}$\,eV and 20\,eV with an upper limit of
42\,eV, respectively.

\begin{figure*}
\begin{center}
	\includegraphics[width=0.8\textwidth]{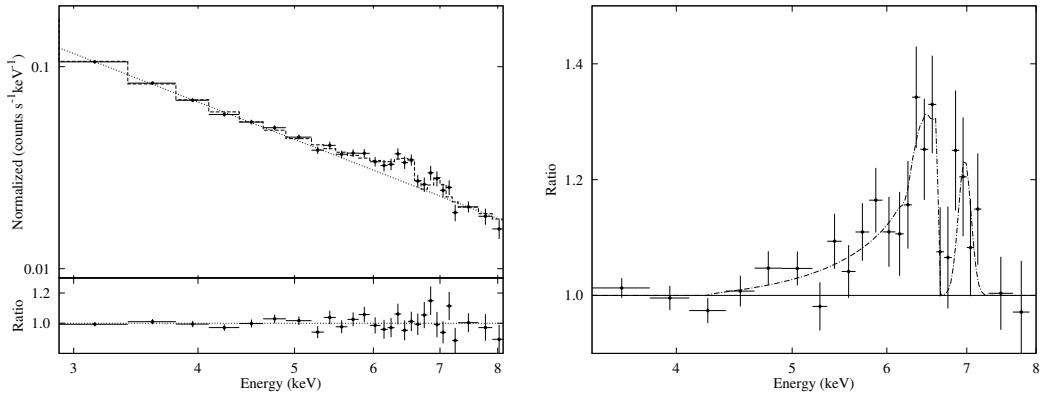}
    \caption{{\it Left-hand Panel:} Spectral fit to the stacked spectrum with
        the best-fit model (dashed line), which comprises a power-law (with all
        parameters free) plus {\sc relline} and a narrow emission line at
        6.97\,keV (Case A).  The residuals as the data to model ratios are
        shown in the lower panel.  {\it Right-hand Panel:} Broad \FeK complex
        of the stacked spectrum shown as the ratio of the data to the best-fit
        continuum (see text).  The best-fit model (as in the left-hand panel;
        dot-dashed line) is shown for a comparison. }
	\label{fig:relline_emi}
\end{center}
\end{figure*}

\begin{figure*}
\begin{center}
	\includegraphics[width=0.8\textwidth]{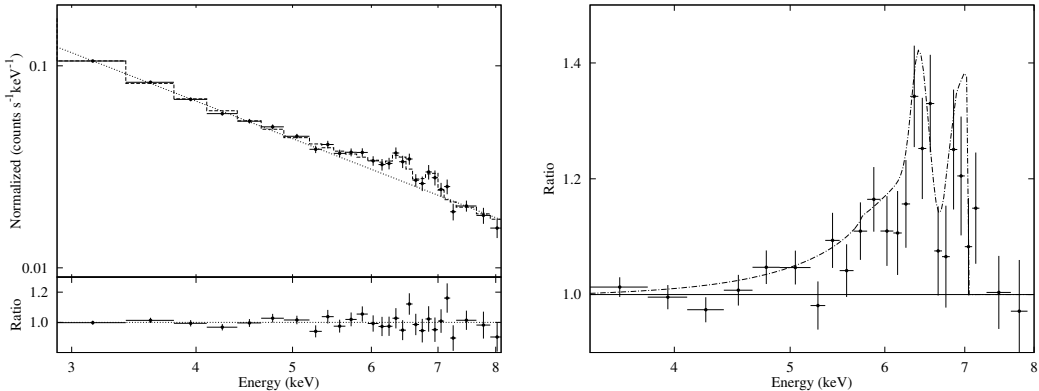}
    \caption{ Same as Figure\,\ref{fig:relline_emi} except that the narrow line
        feature is fitted with an absorption line at 6.67\,keV (Case B) instead of
        an emission line at 6.97\,keV.  }
	\label{fig:relline_abs}
\end{center}
\end{figure*}

\textbf{\textit{Relline model:}} A relativistic \FeK line with high S/N can be
used to constrain the spin of the black hole, if  the inner disc radius extends
to the ISCO, which is a monotonic function of the spin parameter $a$.  So far,
black hole spins have been estimated in about two dozen AGN (e.g.,
\citealt{Reynolds+2013}, including the NLS1 1H0707-495), by using mostly the
relativistic line models (e.g.,{\sc kerrdisk},
\citealt{Brenneman+Reynolds+2006};  {\sc relline}, \citealt{Dauser+al+2010}).
Here we assume that this method can also be applied to NLS1s, as did by others
in previous work to estimate the black hole spin for a few individual NLS1s
\citep{Fabian+al+2009, Marinucci+al+2014}

\begin{figure*}
\begin{center} 
    \includegraphics[width=0.38\textwidth]{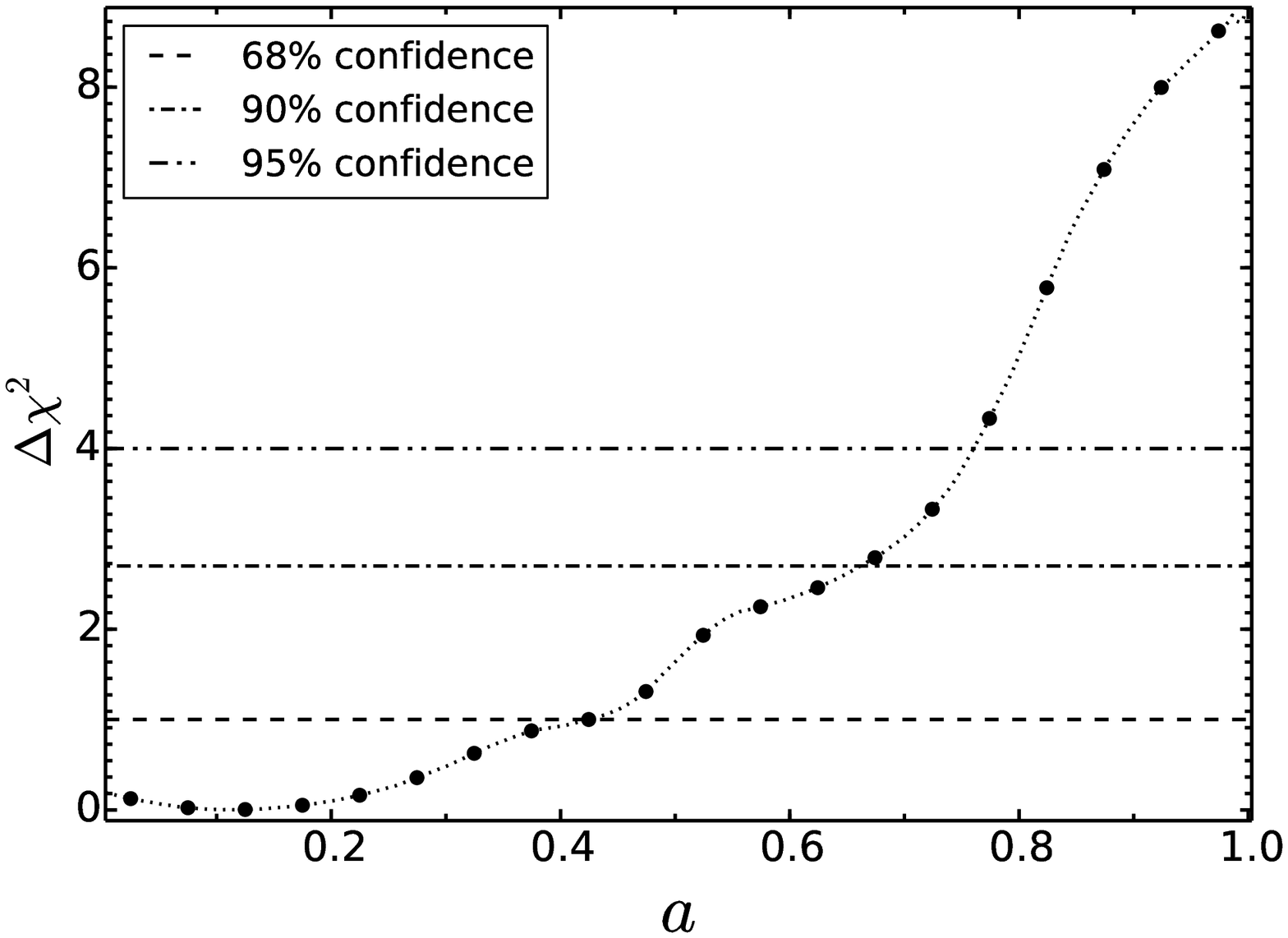}
    \hspace{10mm}
	\includegraphics[width=0.38\textwidth]{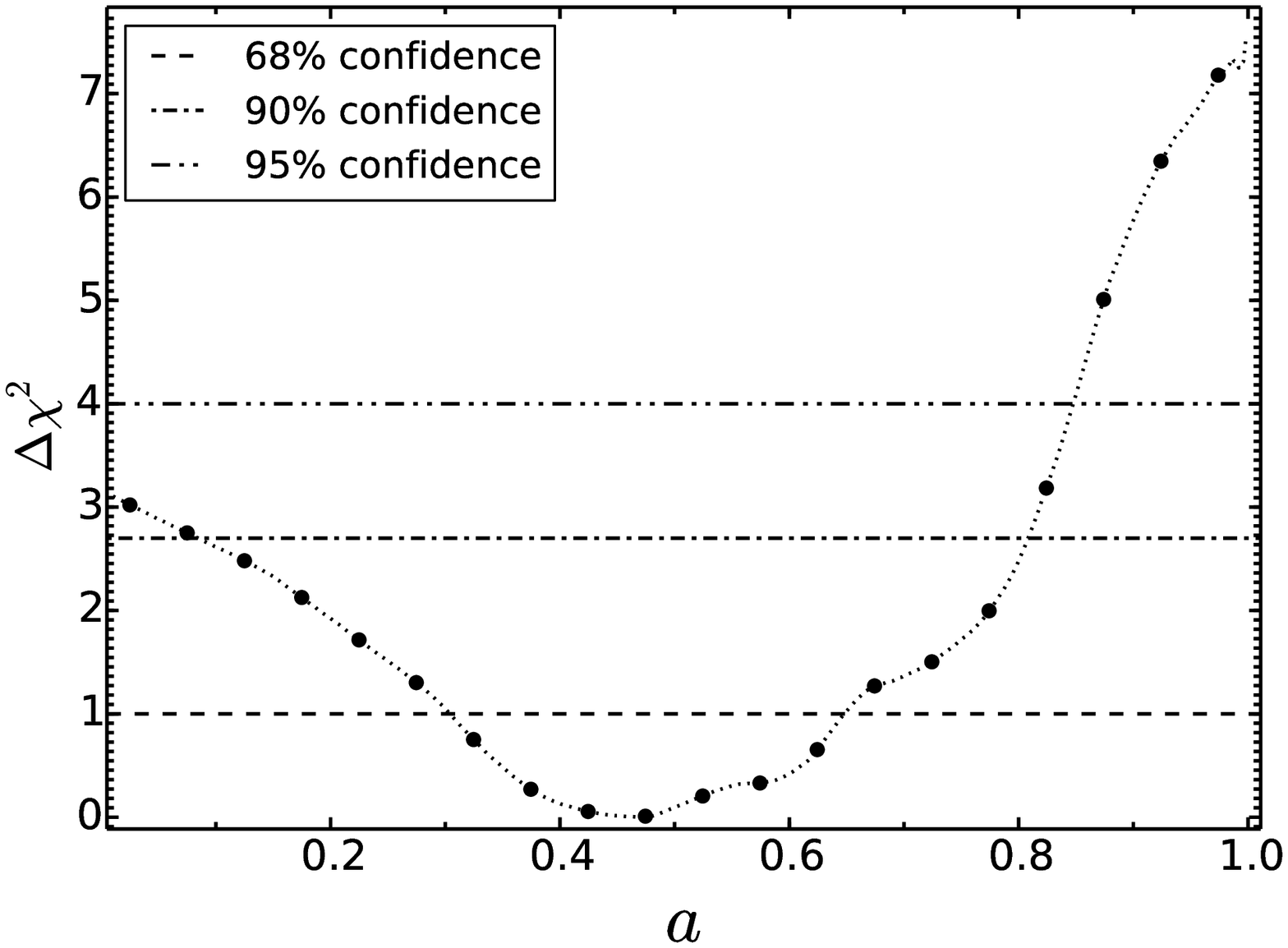}
    \caption{Error contour for the spin parameter, calculated using the {\sc
        relline} model with the inclination angle fixed at 30 degree and the
        emissivity index fixed at 3.0, and the power-law slope
            is fixed at that obtained from
        the stacked simulated continuum. The three horizontal lines, from bottom to top, are
        corresponding to the 68 per cent, 90 per cent and 95 per cent
        confidence, respectively.  {\it Left-hand Panel}: Case A (with a 6.97\,keV
        narrow emission line); {\it Right-hand Panel}: Case B (with a 6.67\,keV narrow
        absorption line).} \label{fig:spin_conf} 
\end{center}
\end{figure*}

We fit the broad-line component with {\sc relline} model. The same as above,
both Case A (with an unresolved 6.97\,keV emission line) and Case B (with an
unresolved 6.67\,keV absorption line) are considered.  Due to the relatively low
S/N of the stacked spectrum, only the broad line energy and the spin of BH are
free parameters. All the other parameters are fixed at default values (e.g.,
the emissivity index is fixed at 3.0 and the disk inclination fixed at 30
degree. Leaving the emissivity index free does not improve the fitting result
and it can not be well constrained.  When left as a free parameter, the
best-fit inclination $i$ is $< 10$\,degree, which is somewhat small for Seyfert 1
galaxies. We thus fix it at 30\,degree).  In Case A, the {\sc relline} model
leads to an acceptable fit ($\mathrm{\chi^2/d.o.f}=23/21$). The broad line
energy is $6.34\pm0.04$\,keV, corresponding to neutral Fe ions.  The spin
parameter is found to be $a<0.66$ (90 per cent significance level,
see Figure\,\ref{fig:spin_conf}, left panel) with a best-fit value of
0.10. The $EW$s of the broad line, narrow \FeK line and narrow
high ionization line are similar to the {\sc laor} model ($EW_{\rm
BL}=364\pm87$\,eV, $EW_{\rm NL6.4~keV}=21$\,eV with an upper limit of
51\,eV and $EW_{\rm NL6.97~keV}=76\pm30$\,eV). In Case B, the model
leads to an equally well fit ($\mathrm{\chi^2/d.o.f}=23/21$). The
fitted energy of the broad line is $6.73\pm0.04$\,keV, implying that the line
may originate from a highly ionized accretion disc. The $EW$s of the broad
component, the narrow component and the absorption component are
$570\pm120$\,eV, $33\pm27$\,eV and $21$\,eV with an upper limit of 43\,eV,
respectively. The best-fit spin parameter is
$0.47^{+0.34}_{-0.39}$ (90 per cent significance level, Figure\,
\ref{fig:spin_conf}, right panel). In either case, the inferred black hole spin
from the composite Fe line profile is not very high.

The above fits are obtained by fixing the power-law index to that obtained from
the stacked simulated power-law spectra. We also elevate this constraint by
leaving the  parameters of the power law component free in the fit. We find
that, in both Case A and Case B, the best-fit photon index is slightly flatter
than, but consistent within errors with, the simulated continuum, i.e.,
$1.92\pm0.08$ for Case A and $1.92\pm0.09$ for Case B. However, the BH spin can
only be loosely constrained: in Case A $a<0.40$ and in Case B
$a<0.84$ at the $68$ per cent confidence level. The equivalent
width of the broad line is $233\pm113$\,eV and $400\pm{227}$\,eV for Case A and
Case B, respectively. The fitted parameters are shown in
Table\,\ref{tab:table3}. The best-fit model for Case A and Case B are shown in
Figures\,\ref{fig:relline_emi} and \ref{fig:relline_abs}, respectively.

\textbf{\textit{Blurred reflection model:}} As a more self-consistent approach,
a reflection component, which is also likely subject to the relativistic
effects, has to be considered. We model the spectrum with the baseline model
plus a blurred reflection model as well as a narrow emission/absorption line.
To implement this, we substitute the {\sc relline} component with a smeared
(cold) disc reflection component, i.e., {\sc pexmon$\ast$relconv}
\citep{Dauser+al+2010}, in Case A. The spin, emissivity index, inclination
angle and normalization are free parameters, while all the other parameters in
the {\sc relconv} component are fixed at default values, i.e., the inner radius
of the accretion disc is fixed at the ISCO of the corresponding spin value; the
outer radius is fixed at $400~R_{\mathrm g}$. The high-energy cut-off in the
{\sc pexmon} component is fixed at 1000\,keV and the solar abundances are
assumed. We obtain an excellent fit ($\mathrm{\chi^2/d.o.f}=17/18$) with the
best-fit photon index of $2.08\pm{0.20}$ and the reflection fraction
$R=1.3^{+1.7}_{-0.7}$. Both the inclination angle (the best-fit
value $i=27$\,degree) and the emissivity index (the best-fit value $q=3.3$) could
not be well constrained and only lower limits are given, $i>22$\,degree and
$q>2.1$. An upper limit on the spin parameter $a<0.35$ (90 per cent
confidence level) could be given if both the inclination and the emissivity
index are fixed at their best-fit values.

The same procedure is applied to Case B but with an ionized reflection (i.e.,
{\sc pexriv} plus a highly ionized emission line with the line energy of
6.67\,keV and the line width of 1\,eV) component. The disc temperature in {\sc
pexriv} is fixed at $10^6$\,K and the inclination angle is fixed at 30 degree.
The disc ionization cannot be constrained at all, and is thus fixed at
$150~\mathrm{erg~cm~s^{-1}}$. Both the emissivity index and the reflection
fraction parameters cannot be effectively constrained. We thus fix the
emissivity index at a range of reasonable values, from 2.4 to 4.0 with a step
of 0.2, and leave the reflection fraction as a free parameter.  We find that
this model can fit the stacked spectrum equally well for different emissivity
indices ($\mathrm{\chi^2/d.o.f}\sim17/18$). The best-fit reflection fraction
$R$ is highly anti-correlated with the emissivity index. Both the fitted photon
index (i.e., from $1.96$ to $2.31$ with a typical 90 per cent uncertainties of
0.35) and inclination (i.e., 24--29\,degree with a typical 90 per cent
uncertainties of 6\,degree) remain consistent within their mutual uncertainties.
The spin is constrained to be $a<0.8$ at the 90 (68) per cent confidence if the
emissivity index is $>3.2$ ($<3.2$).

In conclusion, a detailed examination of the broad emission feature reveals a
highly skewed broad line, which can be well fitted with relativistic line
models of the Fe K$\alpha$. In addition, there is an indication for a possible
narrow feature, either an emission (6.97\,keV, Fe\,XXVI) or an absorption line
(6.67\,keV, Fe\,XXV) superimposing on the broad Fe line. The fitted energy of the
relativistic line is consistent with neutral Fe in Case A, or with ionized Fe
in Case B. In both cases, our analysis using the simple relativistic broad line
model suggests low or intermediate values for the average spins of the black
holes. A more self-consistent modeling incorporating relativistic reflection
tends to be in accordance with these results, although the constraints on the
spin values are of only low statistical significance. Future spectroscopic
observations with higher S/N are needed to better constrain the black hole spin
in these AGNs.

\subsection{Partial covering model}

Although a broad line profile peaked at around 6.4\,keV with a red wing can be
explained by relativistic broad \FeK line which is likely to originated from
the region close to the central black hole. The emergence of a broad profile
may also be due to complex absorption \citep{Miller+al+2008}.
\citet{Gallo+2006} shown that some NLS1s exhibit spectral complexity in their
X-ray spectra and can be explained either by disk reflection or partial
covering \citep[][see also, e.g., Mrk 335, \citealt{Gondoin+al+2002,
Longinotti+al+2007, Grupe+al+2008}]{Tanaka+al+2004}. We thus fit the stacked
spectrum with the partial covering model ({\sc pcfabs} in \Xspec), leaving the
photon index and normalization parameters of the power-law free. In this case,
the partial covering model can fit the stacked spectrum well
($\mathrm{\chi^2/d.o.f}=18/21$) with column density $N_{\rm
H}=4.4^{+2.0}_{-1.2}\times10^{23}~\rm{cm}^{-2}$ and covering factor
$\rm{CF}=0.48^{+0.12}_{-0.15}$. The best-fit photon index
($\Gamma=2.30\pm0.28$) is somewhat larger than, but still marginally consistent
with, the simulated continuum. The $EW$ of the narrow line is about
$\sim43\pm{25}$\,eV. We also try to add an unresolved Gaussian emission line at
6.97\,keV or absorption line at 6.67\,keV to account for the potential emission
or absorption line as shown in Figure \ref{fig:aver_con_sig}. We find that
adding an emission line at 6.97\,keV does not improve the fit at all while the
fitting result is slightly improved ($\mathrm{\chi^2/d.o.f}=17/20$) after
adding an absorption line at 6.67\,keV. The $EW$ of the 6.67\,keV absorption line
is about $\sim$11\,eV (with an upper limit at $\sim$35\,eV).

\section{Discussion}

By stacking the {\it XMM-Newton} spectra of a large sample of NLS1s, we find a
significant broad line profile in the stacked spectrum, which appears to be
elusive in the individual spectra. This broad line profile can be well
explained with the relativistic broad line scenario.  However, it's also
suggested that NLS1s show spectral complexity at X-ray energies
\citep{Gallo+2006} and normally can be explained by either reflection from
accretion disc or absorption by partial covering material along the line of
sight. The broad line emerged in the stacked spectrum can also be fitted with
partial covering model. Partial covering by neutral gas cannot be ruled out as
an alternative explanation to the broad line-like feature shown in Figure
\ref{fig:aver_fit}. Some of the recent results from simultaneously {\it
XMM-Newton} and {\it NuSTAR} observations seem to support the relativistic
scenario \citep{Risaliti+al+2013, Marinucci+al+2014}, however.  We thus take
the interpretation of broad Fe line in this paper, and discuss the implications
of our results in the context of this scenario. Our results thus suggest that
the broad relativistic \FeK line is common in NLS1s, as in other Seyfert
galaxies.  The absence of broad \FeK line in most individual objects with high
S/N in their X-ray spectra has been discussed in \citet{Nandra+al+2007}. In our
sample this may be simply due to the relatively low spectral S/N around the
\FeK line.  This problem is perhaps also aggravated by the relatively steep
continua of NLS1s on average, which means a small number of photon counts at
high energies---thus poorly determined continuum spectra---and in the \FeK
region.

\subsection{Broad Fe line in NLS1s}

Our results suggest that the fraction of broad relativistic \FeK line in NLS1s,
which is still unclear (e.g., $\sim$13 per cent in the full sample of
\citealt{deLaCalle+al+2010} and 2 out of 4 in their flux limited
sample), should not be low.  We notice that the relativistic broad profiles are
perhaps more common in low-luminosity \citep[$L_X <
1\times10^{43}$\,ergs\,s$^{-1}$,][]{Guainazzi+Bianchi+Dovciak+2006} AGN and in
low-luminosity ($L_X < 3\times10^{44}$\,ergs\,s$^{-1}$) or intermediate Eddington
ratio ($\lambda\sim0.1$) for high redshift Type 1 AGN \citep{Iwasawa+al+2012}.
Our results, with a median luminosity of $1.92\times10^{43}$\,ergs\,s$^{-1}$ and
mean luminosity of $6.95\times10^{43}$\,ergs\,s$^{-1}$ in our sample, are in
agreement with these previous work and may support the suggestion that
relativistic broad profiles are more common in AGN with low-luminosities.  Our
results suggest that the broad \FeK line is perhaps also common in AGN with
high Eddington ratio (e.g., NLS1s). This is consistent with the independence of
the detectability of broad line on Eddington ratio \citep{deLaCalle+al+2010}.

The $EW$s of the broad line are $233\pm113$\,eV for Case A and $400\pm227$\,eV
for Case B when fitted with {\sc relline} models, leaving all the parameters in
the power-law component free (see Table\,\ref{tab:table3}). Our results are
consistent with what was found in \citet[][an average $EW$ is of the order of
100\,eV and less than 300\,eV for individual source]{deLaCalle+al+2010} for
low-redshift AGN and is also in agreement with the results by \citet[][an upper
limit with 400\,eV, fitted with the {\sc laor} model]{Corral+al+2008}. The
best-fit energy of the broad line is around 6.7\,keV in Case B, indicating
a highly ionized accretion disc where the line originates from. A high
ionization of thermal origin  may be achieved by slim discs
\citep{Abramowicz+al+1988}.  Alternatively, the high ionization of an
accretion disc may also result from photoionization by hard X-ray
radiation. As suggested in \citet{Ross+Fabian+1993}, the effect of
photoionization becomes important and may produce strong high ionization
emission lines (e.g., Fe\,XXV emission lines) if the accretion rate exceeds
10 per cent of the Eddington rate. Considering the generally high accretion
rates in NLS1s, the  broad \FeK emission observed in the composite
spectrum is more likely arising from highly ionized Fe ions, i.e., from
highly ionized accretion discs, if the observed feature is indeed a 6.7\,keV
absorption line.

\subsection{Neutral narrow Fe K{\boldmath$\alpha$} line in NLS1s}

A neutral narrow \FeK line which may originate from cold material is
ubiquitously found in nearby AGN. The $EW$ of the 6.4\,keV narrow line is
$<60$\,eV, which is slightly weaker than the expected value (e.g., $\sim$60\,eV)
from the empirical Iwasawa-Taniguchi relation \citep[IT
relation,][]{Iwasawa+Taniguchi+1993, Bianchi+al+2007}, but in agreement with
that obtained by \citet{Zhou+Zhang+2010}, in which shown that the $EW$ of
neutral \FeK line in NLS1s is generally weaker than that from BLS1s. The
relatively weaker 6.4\,keV \FeK line may indicate a smaller covering factor of
the reflecting matter which, as pointed out in \citet{Zhou+Zhang+2010}, may be
due to the high radiation pressure \citep{Fabian+al+2008} or outflows probably
associated with the high Eddington ratio accretion in NLS1s
\citep{Komossa+2008}.

\subsection{Ionized narrow Fe K{\boldmath$\alpha$} line in NLS1s}

A narrow emission or absorption line is also suggested in the composite
spectrum at energy around 6.7--7.0\,keV, though the statistical significance
is not very high. We tend to identify it as arising from highly ionized Fe
ions, i.e., Fe\,XXVI in Case A or Fe\,XXV in Case B.  High-ionization
emission/absorption lines, of both low and high velocities, are often detected
in the X-ray spectra of AGN \citep{Bianchi+al+2009, Fukazawa+al+2011}, and also
in sources showing a relativistic broad \FeK line \citep[e.g.,
MCG-6-30-15]{Miniutti+al+2007}.  For example, high-ionization, narrow Fe\,XXVI
emission lines were detected in 21 out of 65 broad line Type 1 AGN and 12 out
of 37 narrow line Type 1 AGN in \citet{Bianchi+al+2009}. Interestingly,
high-ionization emission lines are also found in the stacked spectrum of a
sample of AGN with high accretion rates selected from the XMM-COSMOS field
\citep{Iwasawa+al+2012}. It is thus not surprising to detect such
emission/absorption lines in NLS1s.

In Case A, the $EW$ of the ionized line is $49\pm{32}$\,eV which is in agreement
with the result found in \citet[][5-50\,eV, Seyfert galaxies observed by {\it
Suzaku}]{Fukazawa+al+2011}.  We also notice that \citet{Iwasawa+al+2012} found
a significant highly ionized Fe K line in the XMM-COSMOS Type 1 sub-sample
which includes sources with high Eddington ratios ($\log\lambda>0.6$).
Considering that most NLS1s also have high accretion rate, this may indicate
that there is a potential correlation between the high ionization Fe K emission
line and the Eddignton ratio. Highly ionized Fe emission line can be produced
via recombination and resonant scattering from ionized Compton-thin material
far away from the central region, e.g., broad line region (BLR) or torus, as
shown in \citet{Bianchi+Matt+2002}. The observed $EW$ of $49\pm{32}$\,eV,
assuming a power law index of 2.0 and a solid angle of
$0.26\times4\pi$, can be produced from material with a column density larger
than $\sim10^{23}$\,cm$^{-2}$ and an ionization parameter around
$\log{U_x}\sim0.2$\footnote{The $U_x$ is defined in the energy range 2-10\,keV.
More details can be found in \citet{Bianchi+Matt+2002}}.

In Case B, the absorption line is probably correspond to the Fe\,XXV resonant
absorption line, likely produced in a Compton-thin photoionized material. The
ionization parameter should be low (e.g., $\log{U_x}<0$,
\citealt{Bianchi+al+2005}) since no H-like iron absorption is detected in the
stacked spectrum. The line $EW$ is $18^{+23}_{-18}$\,eV, consistent well with
\citet[][5-40\,eV]{Fukazawa+al+2011}, can be produced by material with column
density larger than $10^{23}$\,cm$^{-2}$.

In any case, an ionized Compton-thin material seems to be required to account
for the potential high ionization emission/absorption feature in the stacked
spectrum. We thus suggest that this kind of material should be ubiquitous in
NLS1s and probably locates far from the central region, e.g., the BLR or torus.

\subsection{Spin of the black holes in NLS1s}

The black hole spin parameter constrained from the relativistic broad Fe
K$\alpha$ line in the stacked spectrum is $a<0.66$ in Case A or $a<0.81$ in
Case B using the {\sc relline} model at the 90 per cent confidence
    level (the statistical significance is somewhat lower if the power-law
continuum slope is set as a free parameter). A higher spin value is also
unlikely to be the case using the blurred reflection model, though with a lower
significance in some cases.  This result indicates that the bulk of the black
holes of NLS1s in the nearby universe (a median redshift $0.085$) either have
averagely low or moderate spins, or distribute in a wide range of spins, from
low spins to high ones; it is ruled out that most NLS1s have extremely large
spins (e.g., $\ga 0.8-0.9$).  This is the first time that the ensemble black
hole spin parameters are constrained for NLS1s as a population. Note that a few
individual NLS1s do have spin measurements with relatively small uncertainties
from high S/N Fe K$\alpha$ line and X-ray reflection continuum, e.g., the spins
of the black holes in two NLS1s, i.e., 1H0707-495 and SWIFT J2127.4+5654, are
estimated to be $>0.89$ \citep{Fabian+al+2009} and $0.58^{+0.11}_{-0.17}$
\citep{Marinucci+al+2014}, respectively. Interestingly, we note that the spin
of J2127.5+5654 is consistent with the ensemble constraints obtained in this
study.

NLS1s may represent an interesting and important phase in the growth and
evolution of supermassive black holes \citep{Komossa+2008}. Given their
relatively small black hole masses and high accretion rates,  there are
suggestions that NLS1s are  at an early stage of AGN evolution
\citep{Mathur+2000}.  The distribution of the spins of the black holes may give
insight into the processes of how their black holes grow and evolve.
Theoretical models have shown that different processes of black hole growth
lead to distinct distributions of the black holes spins. If black holes grow by
merger only, the spin distribution would be bimodal with one peak at
$a\sim$\,0.0 and the other at $a\sim$\,0.7. For processes of merger plus
prolonged accretion, the overall spins are high ($a>0.9$). Low spins ($a<0.1$)
will result from merger plus chaotic accretion processes
\citep{Berti+Volonteri+2008}. There have been suggestions that the evolution of
NLS1s is driven by secular processes instead of major mergers, and their host
galaxies possess mostly pseudo-bulges \citep[][see also
\citealt{Mathur+al+2012}]{OrbandeXivry+al+2011}.

We first consider the prolonged accretion scenario by assuming an initial
non-spinning black hole accreted continuously at the Eddington accretion rate
with a radiation efficiency $\epsilon\sim0.1$.  Since prolonged accretion can
spin up the black hole efficiently, the black hole will reach the maximum spin
after an increase of its mass by a factor of $\sim2.4$ only
\citep{King+Kolb+1999}.  This takes roughly the Salpeter timescale
($\sim5\times10^7$yr, assuming the radiation efficiency $\epsilon\sim0.1$) for
AGN radiating at the Eddington luminosity.  If the black hole spins turn out to
be low or moderate, as is found in our work here, the AGN in NLS1s should be
not only very young, much younger than $<5\times10^7$\,yr, but also with an
initial spin close to 0.  Given the relatively short time left for the black
holes to grow, the current black hole masses of NLS1s would be on the same
order of magnitude as the seed black holes, which would then be
$\sim10^6M_{\sun}-10^7M_{\sun}$.  Since such high-mass seed black holes are not
predicted by the current theories of black hole formation, the pure prolonged
accretion scenario, in which black holes in NLS1s obtained their mass via
continuous gas accretion onto initially much smaller seed black holes, seems to
be inconsistent with our result.

Alternatively, the black holes in NLS1s may grow via mainly the process of
chaotic accretion. Such a process  would lead to low values of the final
average spins, e.g., $a=0.1-0.3$ as found by \citet{King+al+2008}, which is in
good agreement with what we found here ($a<0.66$ in the case where the narrow
feature is an emission line). If confirmed, our result suggests that the growth
and evolution of the black holes in the bulk of NLS1s are likely to proceed via
the chaotic accretion process.

One should also note, however, the black hole spin parameter is constrained to
be widely distributed ($a<0.81$) in the Case B (see Section \ref{sec:subsec4}).
If this is true, then the chaotic accretion cannot be the main process to
control the spin evolution of black holes in NLS1s provided that these black
holes grow up from very small seeds. It is possible that the growth of each of
these black holes is due to chaotic accretion with  many episodes, in each
episode the accretion is prolonged with a significant increase of black hole
mass and significant spin evolution. In order to be compatible with a wide
distribution of spin parameters among NLS1s, the black hole mass increase in
each episode should be less than a factor of few.

To close this section, we note that the standard thin disk model is adopted in
our study to fit the Fe K$\alpha$ line profile for NLS1s as common practice,
however, which may be not valid for some NLS1s. Observations have suggested
that some NLS1s may radiate at luminosities $\ga 30$\,per cent of the Eddington
luminosity, for which a slim disk model may be needed to describe the accretion
\citep{Abramowicz+Fragile+2013}.  In the slim disk model, the disk is puffed up
at the region close to the central black hole, the radiation from this region
may be still efficient or may be not, and some radiation may be even emitted
from the region within the innermost stable circular orbit (ISCO).  Therefore,
the Fe K$\alpha$ emission resulting from a slim disk model, if any, may be
broader or less broad comparing with that expected from a thin disk model. In
this case, it is not easy to directly relate the width and shape of the Fe
K$\alpha$ line emission with the black hole spin due to the complications of
the slim disk model.

\section{Summary}

In this paper, we study the ensemble property of the \FeK line in NLS1s by
stacking the {\it XMM-Newton} spectra of a sample of 51 NLS1s observed in a
total of 68 observations. By virtue of its highly improved S/N, the stacked
spectrum reveals a prominent, broad emission feature, which can be well-fitted
with relativistically broad line models, although a partial
covering absorption model cannot be ruled out. Our result suggests that, as
in Seyfert 1 galaxies, the relativistic \FeK line may in fact be common in
NLS1s, which would be detected in the X-ray spectra of NLS1s with high S/N.
The narrow 6.4\,keV \FeK line is found to be much weaker than the expected
value from the known Iwasawa-Taniguchi relation, which may indicate a smaller
covering factor of the reflection material in NLS1s. The stacked spectrum also
shows a possible high-ionization iron emission or absorption line, produced by
ionised material far away from the central region. We also try to
constrain the average spin of the black holes in NLS1s by modeling the
stacked spectrum with  relativistic line models. Our results tentatively
suggest that the average spins are likely low or intermediate, and
generally very fast spins ($\ga 0.8$) seem to be inconsistent with the
data. This may favour the black hole growth scenario in NLS1s as chaotic
accretion process or chaotic accretion with multi-episodes. Future high
S/N, broad bandpass X-ray observations are needed to confirm these
results.

\section*{Acknowledgement}

The authors thank the anonymous referee for his/her valuable comments which improve the
paper significantly. Z.L. thanks Lijun Gou and Yefei Yuan for useful discussions and suggestions.
This work is supported by the NSFC grants NSF11033007 and 11273027,
the Strategic Priority Research Program ``The Emergence of Cosmological
Structures" of the Chinese Academy of Sciences, grant No. XDB09000000. This
work is based on observations obtained with {\it XMM-Newton}, an ESA science
mission with instruments and contributions directly funded by ESA Member States
and NASA. This research has made use of data obtained from NASA/IPAC
Extragalactic Database (NED) and the Leicester Database and Archive Service at
the Department of Physics and Astronomy, Leicester University, UK.


\bibliographystyle{mn2e}

\label{lastpage}
\end{document}